\documentclass[conference]{IEEEtran}
\IEEEoverridecommandlockouts
\usepackage{cite}
\usepackage{amsmath,amssymb,amsfonts}
\usepackage{algorithmic}
\usepackage{graphicx}
\usepackage{textcomp}
\usepackage{xcolor}
\def\BibTeX{{\rm B\kern-.05em{\sc i\kern-.025em b}\kern-.08em
    T\kern-.1667em\lower.7ex\hbox{E}\kern-.125emX}}
\begin{document}
\bstctlcite{MyBSTcontrol}

\title{Playing With Neuroscience: Past, Present and Future of Neuroimaging and Games}

\author{\IEEEauthorblockN{1\textsuperscript{st} Paolo Burelli}
\IEEEauthorblockA{\textit{Digital Design Department - brAIn lab} \\
\textit{IT University of Copenhagen}\\
Copenhagen, Denmark \\
pabu@itu.dk}
\and
\IEEEauthorblockN{2\textsuperscript{nd} Laurits Dixen}
\IEEEauthorblockA{\textit{Digital Design Department - brAIn lab} \\
\textit{IT University of Copenhagen}\\
Copenhagen, Denmark \\
ldix@itu.dk}
}


\maketitle

\begin{abstract}
Videogames have been a catalyst for advances in many research fields, such as artificial intelligence, human-computer interaction, or virtual reality. Over the years, research in fields such as artificial intelligence has enabled the design of new types of games, while games have often served as a powerful tool for testing and simulation.
Can this also happen with neuroscience? What is the current relationship between neuroscience and games research? what can we expect from the future? In this article, we’ll try to answer these questions, analysing the current state-of-the-art at the crossroads between neuroscience and games and envisioning future directions.
 
\end{abstract}


\section{Introduction}

Throughout the history of artificial intelligence (AI) and game research, the relationship between the two has been mutually beneficial. Games serve as testing grounds for AI algorithms, while AI improves game design, graphics, and player experiences. This synergy helped shape technological advances and creative possibilities~\cite{yannakakis_artificial_2018}. 

A very similar synergy can be observed between machine learning (ML) and neuroscience, the two fields have had mutual inspirations and often shared challenges over the years. Machine learning draws inspiration from the computational principles of the brain -- e.g. by emulating neural networks, AI models learn to recognise patterns and perform complex tasks. In turn, neuroscience benefits from AI as a tool to analyse intricate brain data and develop theories of brain function. Moreover, similar challenges faced by machine learning algorithms and the brain create opportunities for cross-fertilisation~\cite{lindsay_convolutional_2021}.

\begin{figure}
    \centering
    \includegraphics[width=1\linewidth]{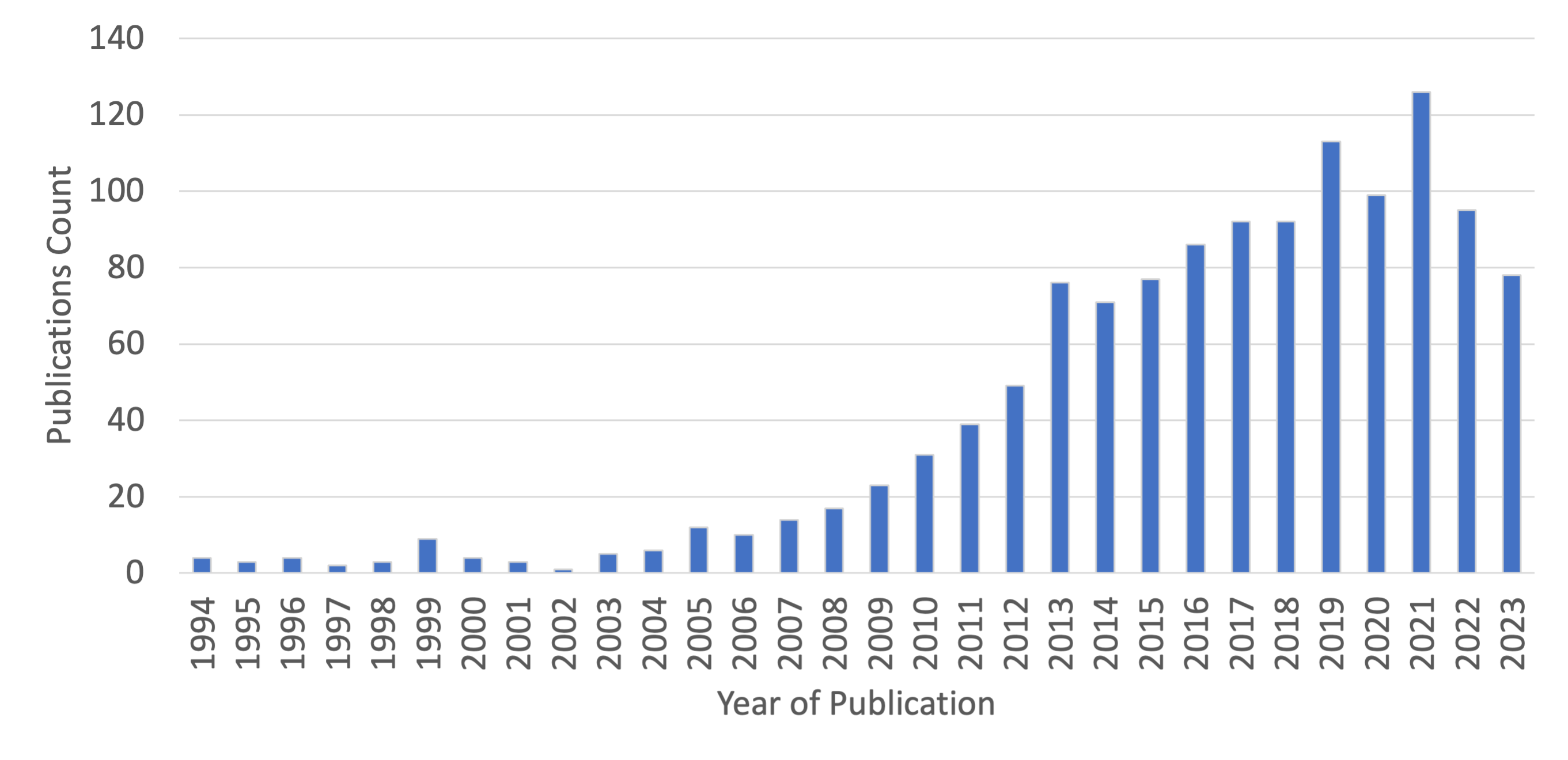}
    \caption{Count of papers published every year between 1990 and 2023 appearing on the Scopus database. The papers have been gathered using the query \textit{``(neuro OR EEG OR fMRI OR MEG OR fNIRS OR BCI) AND (videogame OR "video game" OR "computer game" OR "e-sport" OR "serious game")''}. A total of 1245 papers have been found and used in this visualisation.}
    \label{fig:scopus-pubs}
\end{figure}

In the past decade, we have observed similar synergy emerging between two other fields: neuroscience and game research. As shown in Figure \ref{fig:scopus-pubs}, we found 1245 articles that describe some form of research involving neuroimaging and games on the Scopus database since 1990. This is probably not the full amount; however, it is worth noting that around 90\% of the articles have been published since 2010 and the trend of publication numbers is increasing year by year.

The relationship between these two fields is in part similar to the relationship between AI and games: game research, especially player experience research, can potentially benefit greatly from neuroimaging tools that provide an objective measurement of players' cognitive activity. Some examples of this kind of synergy are studies in which researchers have investigated how EEG and other neuroimaging methods can be used to estimate aspects of player experience, such as challenge~\cite{hegedues_investigating_2023}, engagement~\cite{nunez_castellar_being_2016} or stress~\cite{roy_eeg_2022}.

In addition, games serve as valuable platforms for neuroscience research, providing engaging environments to study cognitive processes and brain activity in different controlled tasks. Researchers leverage games as simulations to explore attention, memory, decision-making, and other mental functions~\cite{yamada_frontal_1998}\cite{schier_changes_2000}. 

These interactive settings allow real-time data collection, allowing analysis of player actions and neural responses. Additionally, games offer opportunities for neurofeedback training, where players learn to regulate brain activity~\cite{delisle-rodriguez_multi-channel_2023}.

Furthermore, as gaming continues to shape our entertainment landscape, understanding its effects becomes paramount. Studies in this area include investigations of the ability of games to improve players' decision-making abilities~\cite{jordan_video_2022} and visual attention~\cite{huang_benefits_2022} as well as identifying potentially harmful effects~\cite{ferrie_video_1994}.

The third and final area of intersection between the two fields we discuss in this article is brain-computer interfacing (BCI). While initially focused on medical applications, BCI research has now expanded its reach to the domain of games and entertainment. This expansion holds immense promise to improve game accessibility and positively impact the lives of individuals with disabilities~\cite{kerous_eeg-based_2018, vasiljevic_braincomputer_2020, marshall_games_2013}.

In the context of games, BCIs offer several advantages. First, they allow users to adapt and control games using brain signals in addition to traditional physical and mental abilities. For ``abled'' users, this means exploring new ways to interact with games beyond conventional input devices such as keyboards or controllers~\cite{bos_human-computer_2010}. Second, BCI technology enables non-intrusive communication between the brain and the game environment. Users can manipulate game elements, navigate virtual worlds, and even achieve in-game objectives by harnessing their brain activity. This approach is particularly valuable for people with physical limitations, such as those who cannot use their hands or limbs effectively~\cite{aguado-delgado_accessibility_2020}. Third, BCI research aims at real-time adaptation—monitoring cognitive states and adjusting game parameters accordingly. Imagine a game that dynamically responds to a player’s focus, attention, or emotional state, creating a personalised and engaging experience~\cite{yannakakis_affective_2023}.

Furthermore, games can serve as a powerful platform for developing and refining BCI applications. Unlike clinical settings, games provide an environment in which users willingly engage for entertainment. This engagement fosters motivation and active participation, crucial for successful BCI training. In addition, games offer a diverse range of scenarios, ranging from high-stakes battles to serene puzzles, that allow researchers to explore various BCI paradigms. As BCI technology matures, it has the potential to transform lives~\cite{pinheiro_wheelchair_2016}.

All of the aforementioned research areas come with challenges such as ethical considerations, data privacy, and the need for robust calibration. Yet, the potentials are vast: personalised gaming experiences, therapeutic interventions, and inclusive design, among many others.
In the forthcoming sections, we delve deeper into these areas, describing the state-of-the-art, isolating the existing open challenges and identifying potential opportunities and future directions.

\section{A Primer on Functional Neuroimaging}

Functional neuroimaging is the discipline of recording brain activity with the purpose of \textit{understanding} some cognitive process. The brain consists of a large web of cells called neurons ($\sim$80 billion), that communicate by sending chemical neurotransmitters. The signal is carried through the neuron as an electrical charge, which is released at connection points called synapses. The neocortex is the outermost layer of the brain, it is very dense in neurons, and most of the uniquely human cognition is thought to happen in the neocortex (cortex/cortical for short). Therefore most functional neuroimaging techniques focus on measuring cortical activity. The two most common techniques are electroencephalography (EEG)~\cite{teplan_fundamentals_2002} and functional magnetic resonance imaging (fMRI)~\cite{logothetis_what_2008}. These two techniques are complementary in nature, providing their own opportunities and challenges. 

\begin{figure}
    \centering
    \includegraphics[width=1\linewidth]{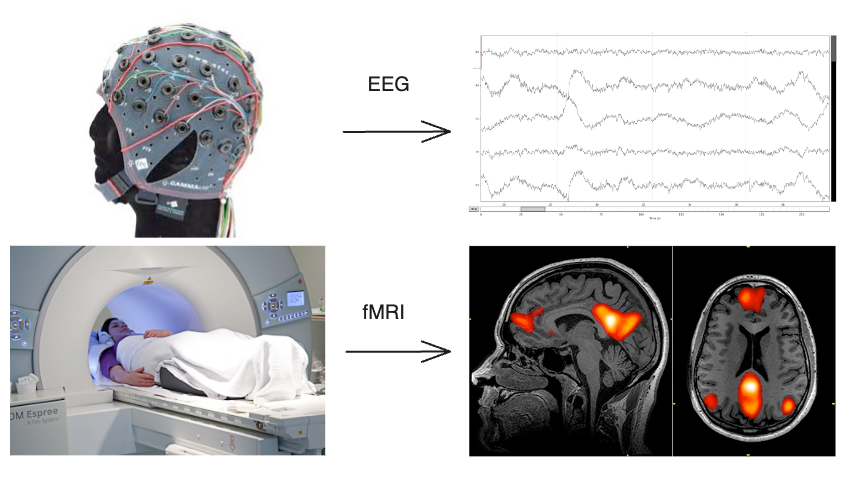}
    \caption{A graphic of how EEG and fMRI is recorded and an example of a standard visualisation. Image is adapted from: g.tec (EEG cap), Hvidovre Hospital Radiologisk Sektion (MR scanner), The Harvard Gazette (fMRI visulisation)}
    \label{fig:neuroimaging}
\end{figure}

EEG is the most common imaging technique to be applied in settings outside of explicit neuroscientific research - e.g. in consumer research~\cite{lin_applying_2018}. This is because EEG generally has low cost, high degree of freedom in experimental design, has a large literature~\cite{kappenman_oxford_2011}\cite{gable_oxford_2022} and easy to use software tools for analysis~\cite{delorme_independent_2012}. EEG utilises electrodes on the scalp to measure electrical potentials generated by neural firing, a standard EEG system has between 8 and 64 electrodes simultaneously recording (see Fig. \ref{fig:neuroimaging}). EEG measures large, synchronised cortical activity. A common metaphor to explain this concept is to say that an electrode on the scalp is like a microphone at a football stadium. We are not able to hear individual conversations, since they are all talking over each other. However, when the audience expresses something in synchrony, we pick it up easily. This is why a large portion of EEG research is focused on 'brain waves', large oscillations in neural activity which are constantly created in the brain. In the metaphor, this would be the audience chanting together. Brain waves are divided into bands, typically measured by Fourier analysis. In general, lower frequency bands (e.g. $\alpha$: 8-13 Hz or $\beta$: 13-30 Hz) signify inhibition of a brain area, whereas higher frequency waves (e.g. $\gamma$: 30-100 Hz) indicate activation. The metaphor also explains how EEG has very poor spatial resolution since the sensor number is very low compared to a number of neurons, additionally, the skull is not very conductive causing a smearing effect of the signal across the scalp. However, EEG can sample very quickly ($\geq$1000 Hz) giving a very high temporal resolution, making it ideal to study phenomena where precise timing is important (reactions) or where it unfolds over time (dynamics). Typically, reactions are studied in the form of event-related potentials (ERPs), where one looks at the brain's immediate reaction to different stimuli or events. In the stadium metaphor, this would be when the audience all reacts to a specific event simultaneously, we can hear it in the microphone. All in all, EEG is an excellent tool to study cognition, but one needs to be aware of the limitations as described here.
fMRI is the most common functional imaging technique within cognitive neuroscience because of its ease of interpretation~\cite{friston_statistical_1994}, high spatial resolution and high signal-to-noise ratio. The main drawback is that one needs access to an MR scanner of $\geq$1.5 T strength. fMRI measures the blood flow in response to neuron activity. This means fMRI is limited by the speed of the vascular system, resulting in a typical protocol taking a full-brain image every 2 seconds, giving very poor temporal resolution. The MR scanner takes 3D images and is therefore not limited to cortical activity alone - fMRI is able to scan sub-cortical structures like the limbic system e.g. hippocampus (memory) and amygdala (emotion). 3D images of the brain are generated, which can be easily visually inspected and analysis is in general straightforward, comparing how active different areas were during tasks (see Fig. \ref{fig:neuroimaging}). However, since the MR scanner needs a strong magnetic field to work, the subject needs to lie down and any potentially magnetic material (including standard wiring) cannot enter the vicinity of the scanner. Furthermore, movement generates a lot of noise in the signal, which limits the possible experimental designs. In general, playing games is challenging on the MR scanner, although there are exciting advancements that expand the type of controllers one can give subjects during scanning~\cite{harel_gamer_2024}.

\section{Games as a Research Platform}

One of the first intersections between games and neuroscience explored by researchers is the study of video games and their impact on brain function, behaviour, and overall cognitive performance. From the first work in the area, researchers have investigated the possible impacts of games on children and young adults' health, such as epilepsy~\cite{badinand-hubert_epilepsies_1998}, mental health~\cite{aliyari_effects_2015} and sleep patterns~\cite{peracchia_exposure_2018}.

Alongside the potential negative effects, several studies have also investigated the potential positive impacts of games on cognitive performance and emotional behaviour. For example, Russoniello et al.~\cite{russoniello_eeg_2009} investigated the effects of playing the popular casual video game Bejeweled II on mood, stress and physiological responses and, in their study, found that the changes in EEG and heart rate observed after playing Bejeweled II were consistent with improved mood and relaxation.

Other researchers found evidence of cognitive improvements: Mondejar et al.~\cite{mondejar_correlation_2016} explored how video game mechanics impact cognitive activation and found evidence suggesting that playing action games helps the development of executive functions. Similarly, Wan et al.~\cite{wan_measuring_2021} have explored the relationship between virtual reality (VR) games and cognitive ability, finding that VR positively impacts the activation of working memory compared to standard three-dimensional games. 

Bae et al.~\cite{bae_effects_2016} conducted an experiment to assess the impact of gameplay activities on children with intellectual disabilities. Their study found that, after participating in-game activities, the experimental group showed significant improvements in attention-related EEG waves, suggesting improved cognitive functioning in terms of attention and focus. Furthermore, the same experimental group demonstrated improvements in social skills after the intervention and also showed better self-control after gameplay sessions.

The observation that certain games, especially those that involve strategy, problem solving, or memory, can improve cognitive functions has emerged in multiple other studies~\cite{jordan_video_2022,huang_benefits_2022} highlighting the potential of serious games to be used as a training tool to improve them.

Shenjie et al.~\cite{shenjie_two_2014} explored the impact of a neurofeedback-based brain-computer interface game on enhancing attention and cognitive skills in healthy individuals: In the proposed training paradigm, participants played the neurofeedback game regularly over a period of 5 days. The game required players to memorise a set of numbers in a matrix and correctly fill the matrix using their attention and the results showed that the neurofeedback game had a positive impact on attention enhancement and cognitive skills. Similar results were reported by Thomas et al.~\cite{thomas_enhancement_2013}, Ballesteros et al.~\cite{ballesteros_effects_2017}, by Alchalcabi et al.~\cite{alchalabi_focus_2018} and by Israsena et al.~\cite{israsena_brain_2021} on older individuals.

Another potential contribution of games as a platform for neuroscience has been highlighted in studies such as~\cite{alchalabi_focus_2018} and~\cite{mcwilliams_feasibility_2021} that investigate the application of games as diagnostic tools for cognitive assessment.

\section{Brain-Computer Interfaces}
An area of research that already has a flourishing interdisciplinary community is the field of brain-computer interface (BCI). BCI systems aim to let a user control a computer with their brain, without physical movement. They do this by recording brain activity through some cognitive neuroimaging technique and applying a decoder model to infer intent. BCI systems have important clinical applications including prosthetic devices~\cite{muller-putz_eeg-based_2005}, virtual keyboards~\cite{birbaumer_spelling_1999} and motor recovery training~\cite{jamil_noninvasive_2021}. The most common imaging technique here is EEG, as it offers high temporal resolution and minimal constraints on the user (see section above). Typically, the decoder model is a machine learning model, which must be trained specifically for each user, requiring many training samples before good performance is achieved. Here, games play an important role, as they offer a uniquely fun and safe interactive environment in which one can train and test their BCI systems. BCI has had a long-lasting community involving researchers from neuroscience, AI, games and engineering~\cite{marshall_games_2013, islam_editorial_2023}. 

\subsection{BCI for games}
For a detailed review of the current state-of-the-art techniques in BCI, we refer to other more specialised reviews, e.g.~\cite{peksa_state---art_2023} and~\cite{varbu_past_2022}. Here we provide a brief overview of common techniques and applications related to games and gameplay. There are three common frameworks for BCI systems: P300, motor imagery (MI), and steady-state visual evoked potentials (SSVEP)~\cite{kerous_eeg-based_2018}. P300 is an ERP component that happens at 300 ms after stimulus and is modulated by attention. By intentionally bringing attention towards one of multiple objects on the screen, the system can decode the intent of the user. P300 is a reliable measure, but requires a long training procedure and is inherently slow due to the 300 ms delay. It is often used in clinical settings, and has been used in assisting in identifying Alzheimer's disease by a board game BCI system~\cite{mcwilliams_feasibility_2021} and cooperative games between able-bodied and people with severe disabilities~\cite{rybarczyk_cooperative_2019}. SSVEP exploits how different flicker frequencies create similar frequency patterns in brain signals. Multiple objects on the screen can be shown with different flickering frequencies and the one looked at should give a strong signal at that frequency. Because frequencies can be very high, there is no built-in delay, as in the P300 paradigm, however, SSVEP can be sensitive to noise from other light sources and requires high mental engagement. SSVEP has also been used in 3D and VR games with some success~\cite{koo_immersive_2015, lotte_brain-computer_2011}. Because both P300 and SSVEP require sustained attention, games have been developed as attention training platforms for people suffering from Attention-Deficit Hyperactive Disorder (ADHD)~\cite{thomas_enhancement_2013, ali_3d_2015}. The last popular BCI technique is motor imagery, where the player is asked to imagine a movement, but not performing it. This creates a strong signal in the part of the motor cortex related to a target body part. Popular movement choices are left/right arm, legs and tongue~\cite{coyle_eeg-based_2011}~\cite{yang_development_2018}. This has the upside of coupling control of in-game events with control of body parts, making the gameplay experience more natural. However, measures can be unreliable and needs long training periods to work with high accuracy. Motor imagery games has been used to give physically disabled people access to games, which would not be possible with traditional control systems~\cite{pinheiro_wheelchair_2016, kauhanen_eeg-based_2007}.
A main limiting factor in all these systems is the low number of available choices, requiring designers to work around this when designing the interface. General improvement in the performance of various aspects of the system is also still needed. Training time is necessary for each player for decent performance, which could be shortened by having better decoding models. A major research focus has been to make BCI work on consumer-grade equipment~\cite{vasiljevic_braincomputer_2020}~\cite{amprimo_measuring_2023}~\cite{van_vliet_designing_2012}, which would greatly enhance the adoption possibilities. While BCI still needs major developments before it has widespread adoption, it offers an exciting research area with an already existing interdisciplinary community, with the potential to make life-changing products for some people.

\section{Player Experience Research}

Developing accurate and reliable estimations of player experience (PX) is one of the main challenges within the game research community~\cite{wiemeyer_player_2016}; Providing high-quality experience estimates is vital for game companies to evaluate their products~\cite{drachen_games_2018} and enable personalised game experiences~\cite{pedersen_modeling_2010}.

Questionnaires such as PXI~\cite{abeele_development_2020} and PENS~\cite{ryan_motivational_2006} offer validated and reliable operationalisations of several aspects of the PX; however, they are limited in describing overall experiences not connected to specific game events.

Several studies have investigated and proposed ways to perform the same measures in real time using behavioural data from the game~\cite{melhart_your_2019, drachen_behavioral_2015} and from the behaviour of the physical player~\cite{burelli_non-invasive_2014}.
However, these metrics provide insights into player behaviour; they are limited in their ability to capture emotional signals. 

Physiological measures, such as skin conductance level, heart rate, and brain activity reveal covert changes in a player’s state. These metrics have shown the potential to evaluate emotional responses that are otherwise hidden from the naked eye~\cite{yannakakis_psychophysiology_2016, robinson_lets_2020}.

One of the most promising physiological signals investigated is EEG since it is a direct correlate of the players' cognitive activity. Studies have investigated the operationalising of multiple aspects of the player experience, including flow~\cite{berta_electroencephalogram_2013}, challenge~\cite{hegedues_investigating_2023}, emotions~\cite{kollias_affective_2014}, engagement~\cite{mcmahan_evaluating_2015} and stress~\cite{roy_eeg_2022}.
Researchers have explored the adoption of EEG-based operationalisations both as an off-line tool for game user research~\cite{wehbe_introduction_2014} and automated game personalisation~\cite{stein_eeg-triggered_2018}.

Although these studies offer insightful results, the field still lacks widespread validation and standardisation, especially when compared to PX estimation instruments, such as questionnaires. In an attempt to start to address these limitations, Alakus et al.~\cite{alakus_database_2020} have recently published a dataset containing electroencephalography-based data for emotion recognition. 

EEG signals were collected from 28 different subjects with a portable EEG device called the 14-channel EMOTIV EPOC+. The participants played four different computer games that captured emotions (boring, calm, horror, and funny) for 5 minutes, resulting in a total of 20 minutes of EEG data available for each subject. Subjects also rated each computer game based on the scale of arousal and valence using the Self-Assessment Manikin (SAM) form. 

The main limitation of this work is the quality of the hardware used for the data collection, as the EMOTIVE EPOC+ has limited capabilities in terms of sampling frequency and measurement accuracy~\cite{duvinage_performance_2013}. However, along with other works such as~\cite{melhart_arousal_2022}, this research work is an important first attempt in the direction of creating standardised datasets and baselines that can help the field progress.

\section{Challenges, Opportunities and Future Directions}

The state-of-the-art review described in the previous sections of this article showcases the existence of an active field of research that intersects games and neuroscience in multiple dimensions.

As in many other interdisciplinary fields, the main challenge stems from the multiple different skills required to be able to conduct valid research. Researchers working in this come from varied backgrounds including, among others, games design, computer science, psychology, and neuroscience that rely on different epistemologies and methodologies. Therefore, for the future success of this kind of field, we believe that efforts to facilitate communication and create spaces for collaboration are necessary.

Beyond these typically interdisciplinary challenges, the field has a number of specific challenges and potentials, and, in this section, we will try to highlight the current main challenges in the field, suggest future direction of research, and analyse the potential impact.

\subsection{Technology and Applicability}

The most obvious obstacles to the widespread adoption of EEG and fMRI-based research outside of medical applications are the cost and complexity of operating the hardware. Although there is increasingly less expensive EEG recording equipment available on the market, high-quality research-orientated EEG amplifiers still have unaffordable prices for individual researchers.
The high cost of quality hardware does not only reduce the viability of the research but also its applicability as products such as adaptive games based on EEG remain a prerogative of laboratory experiments. 

As confirmed by the advent of cheaper albeit lower-quality hardware, this problem is likely going to be limited to this early era of research between neuroscience and games. However, together with the complexity of collecting quality EEG and fMRI data, the current cost of the hardware is a hindrance to the current potential development of the field. 

A potential solution to this problem is to make more efficient and effective use of the data that is being recorded but sharing it and developing standard datasets.

\subsection{Standard Datasets}

A major resource for establishing connections across domains is to create standardised datasets that can be used by researchers from both fields. Well-documented and open datasets have become increasingly popular in the neuroscience community, with large-scale datasets and benchmarks being open to the broader research community~\cite{grootswagers_human_2022,jayaram_moabb_2018}. 
These serve to allow researchers from the AI community to be able to work on problems in neuroscience, exploiting insights from specialists from other fields. We propose that game studies using neuroimaging should follow the same path, aiming to create datasets that are accessible and useful for both neuroscience and game research. 

As discussed below, games allow for dynamic testing environments without loss of information. Therefore, we highly suggest researchers to log all relevant information presented to the subject during recording, making the dataset richer and deeper. This way, the dataset can be reused by other researchers asking novel research questions not addressed by the original experimenters. 

This entails keeping track of as many game events, player inputs, visual, and auditory stimuli as possible, making the gameplay experience ideally fully reconstructable. Similarly, raw neuroimaging data should also be conserved and available. While feature extraction and preprocessing are common strategies for analysis, other researchers might be able to make different strategies work. Raw data is the format which is maximally flexible for repurposing datasets. Although, here it must be mentioned that explicit and informed consent must be gathered from subjects to publish and use their data for scientific purposes. We expand on this point in a later section.

Some early attempts have been made to produce such datasets~\cite{duvinage_performance_2013}; however, we believe that more is needed.
We recommend following the Brain Imaging Data Structure (BIDS)~\cite{gorgolewski_brain_2016}. Lastly, annotating and documenting the dataset should be of high priority. Events should be clearly labelled, and recommended preprocessing, feature extraction and analysis pipelines should be clearly described. This would allow other researchers to build on existing work more easily. Data is a valuable resource and should be shared openly, making non-experts able to expand on its use. 

\subsection{Ecological Validity}

Games have the potential to greatly improve the protocols used in cognitive neuroscience, both to study cognitive phenomena within games and phenomena that potentially translate beyond games and are of interest to neuroscientists, psychologists, social scientists and other researchers using psychological experiments. 

To ensure the reliability, validity, and allow the researchers to reasonably look for patterns in the data, the standard experimental procedures in neuroscience require the participant to engage in mostly simple tasks and stimuli, repeated multiple times, often with minimal variations. This is done with the purpose of isolating the phenomenon of interest, making sure the differences between conditions can be isolated in the recorded data.

However, there are three major drawbacks of this approach:
\begin{itemize}
    \item the rigidity of the experiment makes the setting artificial reducing ecological validity and, therefore, the ability to generalise results to the real world,
    \item the static nature of the experiment makes an assumption of a constant stereotypical behaviour to each repeated stimulus, missing out on potential complexity in how contexts, environments and duration might affect the cognitive processes.
    \item the experiment is potentially very repetitive and the subject cannot sustain engagement in the task for longer periods of time, risking poor data quality and limiting the length of recording sessions.
\end{itemize}

We argue that games have the potential to solve all three drawbacks and, if implemented correctly, while still keeping the desired properties of the original design. 

Games are explicitly designed to be engaging; therefore, experimental designs that draw on game design could greatly increase subject engagement. This would allow longer experiments with higher data quality. The engaging nature of games also make the generalisations to real-world scenarios easier. Subjects immersed in the world of the game are more likely to forget they are in a laboratory participating in an experiment, making their behaviour more natural and therefore more ecologically valid. 

The dynamic nature of games allows subjects to perform similar tasks in different settings, making it possible to study the same phenomenon in varied contexts. For example, the same action is naturally repeated in different areas of the game with different visual and auditory stimuli, allowing the isolation of the cognitive process from the perceptual one on vice-versa. Furthermore, all this can be achieved while keeping track of the environment, automatically creating richly annotated datasets of neurological behaviour. 

These characteristics make cognitive neuroscientific experiments dynamic, interactive, and more engaging; and speak to the idea that modern neuroscience needs a new guiding assumption of the brain by acknowledging the complexity of the underlying mechanisms~\cite{westlin_improving_2023}, which current experimental methods do not do. All of this suggests that including games in experimental protocols would add much value to the neuroscientific community.

However, besides the reductionistic nature of the current common experimental protocols, one of the reasons for their simplicty is also the difficulty in interpreting complex non-linear patterns that would emerge from the interaction with digital artifacts such as games. 

EEG recording in a game session can produce an enormous amount of data describing the players' actions, the game state and the psychophysiological signals. This contrasts with the relatively small amount of triggers that are usually used to label EEG data in classical experiments and creates a new challenge in terms of data processing and analysis. Techniques such as averaging over repetitions and statistical hypothesis testing become insufficient.
However, the contemporary rise of deep learning and other advanced machine learning methods offers a potential solution to both real-time processing and post-hoc data analysis in these scenarios.

\subsection{Deep Learning, Neuroscience and Games}

The last years have seen a massive surge of research in deep learning techniques. With deep learning (DL) models being loosely based on principles from neuroscience, the interplay between these two fields has come into more focus~\cite{marblestone_toward_2016}. Models are now performing at human or even super-human level in some isolated cognitive tasks, that normally were only possible to study through biological systems. 
Cognitive phenomena are often complexly integrated in the brain and may vary greatly from setting to setting and person to person. DL models have shown to be capable of capturing these nonlinear relationships without losing generality, and harnessing this power provides great opportunities in neuroscience~\cite{kriegeskorte_deep_2015}. Conversely, inspiration for improvements in model performance on artificial intelligence tasks is often found in how human or animal systems solve them~\cite{lindsay_convolutional_2021}. Simple vertebrae like fish are able to solve tasks that, at the moment this article has been written, even the best DL models are unable to, like one-shot learning and integrated 3-D rotations of objects. 

Researchers working in either field should consider how insights from their counterparts could aid them. We argue here that games is well positioned to act as a mediating middle point between AI and neuroscience research. Game research is already highly integrated with AI~\cite{yannakakis_artificial_2018}, and we think that the overlap between the three fields could create novel research opportunities with new potential synergies.
    



\subsection{Personal data and privacy concerns}

While open data sharing is essential in a research environment, the same cannot be said when data is recorded for commercial use. When dealing with physiological data, users are at risk of sharing extremely private information, including data that could be used for reasons that go far beyond the intended examples aforementioned in this article. 
The risk of unwillingly exposing private clinical information about users increases. EEG can be used to diagnose epilepsy, so exposing subjects to potentially trigger stimuli and recording their EEG might reveal diagnosed or undiagnosed epilepsy. 
The responsibility of managing these risks should be on the side of the developer and researcher, and this must be thought of early in any design process involving collecting physiological data. 

A promising research direction, that addresses some of these concerns, approaches physiological data collection and processing in two: EEG data is pre-processed locally by encoding into a common anonymised latent space, data represented in this latent space is then stored or used for any machine learning task~\cite{bethge_domain-invariant_2022}. This approach has the potential to both mitigate the risk of personal data sharing and of creating more personalised private models that address people's neural diversity.



\subsection{Conclusion}

The article proposes an in-depth overview of the intersection between neuroscience and game research. We identify three main areas of research -- \textit{games as a neuroscience research platform}, \textit{neural player experience research} and \textit{brain-computer interfacing} -- we discuss their state and highlight potential future directions.

In particular, we argue that games have the potential to be the key to more ecologically valid neuroscientific research, and at the same time, neuroimaging and neuroscientific models have the potential to revolutionise player experience research. However, for these two predictions to come to fruition, we see machine learning having an important role. By leveraging deep learning, we can potentially create more powerful and accurate models of the mind. This would allow processing of the large amount of data created in game-based experiments and developing more accurate and robust models of players' affect and cognition. 

We conclude by highlighting the potential risks that large-scale neuroimaging data collection and processing for players' privacy, while encouraging more research on methods to mitigate these risks.

\bibliographystyle{IEEEtran}
\bibliography{IEEEabrv, nourl, references}

\end{document}